\newcommand{\paperOption}{conference,a4paper}
\newcommand{\useOtherPackages}{%
 \usepackage[pdfstartview=FitH,hidelinks]{hyperref}%
}
\newcommand{\omitstyle}{\color[rgb]{0.25,0,0.75}}
\newcommand{\omitted}[2][]{{\omitstyle#2}}
\newcommand{\reducevspace}{\omitted[\vspace{-1.9ex}]{}}

\InputIfFileExists{option.aux}{%
 \typeout{Compiling options overridden!}%
}{}

\documentclass[\paperOption]{IEEEtran}

\usepackage{amsmath}
\usepackage[noadjust]{cite}
\usepackage{graphicx}
\usepackage{color}

\useOtherPackages

\newcommand{\wholeline}[1]{\IEEEeqnarraymulticol{3}{l}{#1}}

\newtheorem{theorem}{Theorem}[section]

\newtheorem{lemma}[theorem]{Lemma}
\newtheorem{proposition}[theorem]{Proposition}
\newtheorem{remark}[theorem]{Remark}

\newcommand{\borel}{\mathcal{B}}
\newcommand{\diff}{\mathrm{d}}
\newcommand{\distribution}{\mathcal{P}}
\newcommand{\eqdef}{:=}
\newcommand{\kl}[2]{\mathrm{D}(#1\|#2)}
\newcommand{\klbig}[2]{\mathrm{D}{\left(#1\Big\|#2\right)}}
\newcommand{\monotone}{\mathcal{M}}
\newcommand{\nb}{\mathrm{e}}
\newcommand{\norm}[1]{\left\|#1\right\|}
\newcommand{\real}{\mathbf{R}}
\newcommand{\set}[2]{\{#1\colon #2\}}
\newcommand{\support}{\mathrm{supp}}
\newcommand{\useless}{\mathrm{U}}

\begin{document}

\title{A Geometric Property of Relative Entropy and the Universal Threshold Phenomenon for Binary-Input Channels with Noisy State Information at the Encoder}
\author{%
 \IEEEauthorblockN{%
  Shengtian Yang\IEEEauthorrefmark{1}\IEEEauthorrefmark{2}
  and Jun Chen\IEEEauthorrefmark{2}}%
 \IEEEauthorblockA{%
  \IEEEauthorrefmark{1}%
  School of Information and Electronic Engineering, Zhejiang Gongshang University, Hangzhou 310018, China}%
 \IEEEauthorblockA{%
  \IEEEauthorrefmark{2}%
  Department of Electrical and Computer Engineering, McMaster University, Hamilton, ON L8S 4K1, Canada}%
 \IEEEauthorblockA{%
  Email: yangst@codlab.net, junchen@ece.mcmaster.ca}%
}

\maketitle

\begin{abstract}
Tight lower and upper bounds on the ratio of relative entropies of two probability distributions with respect to a common third one are established, where the three distributions are collinear in the standard $(n-1)$-simplex.
These bounds are leveraged to analyze the capacity of an arbitrary binary-input channel with noisy causal state information (provided by a side channel) at the encoder and perfect state information at the decoder, and in particular to determine the exact universal threshold on the noise measure of the side channel, above which the capacity is the same as that with no encoder side information.
\omitted[%

\color{red}%
\textnormal{The extended version including omitted proofs will be available at \url{http://arxiv.org} or via \url{http://www.yangst.codlab.net}.}%
]{}
\end{abstract}

\IEEEpeerreviewmaketitle

\section{Introduction}

It is shown in \cite[Lemma~1]{xu_when_2017} that for any binary-input channel with  noisy causal state information (provided by a side channel) at the encoder and perfect state information at the decoder, if the side channel is a generalized erasure channel and the erasure probability is greater than or equal to $1-\nb^{-1}$, then the capacity is the same as that with no encoder side information.
In other words, $1-\nb^{-1}$ is a universal upper bound on the erasure probability threshold, which does not depend on the characteristics of the binary-input channel and the state distribution.
However, as is noted in \cite[Footnote~2]{xu_when_2017}, this bound is not tight, so determining the exact universal erasure probability threshold remains an interesting open problem.
It is worth mentioning that, with the erasure probability replaced by a suitably defined noise measure, this universal threshold holds for all side channels (see \cite[Theorem~3]{xu_when_2017} and \eqref{gamma.definition}).

We shall settle this open problem by characterizing a certain geometric property of relative entropy (also called the Kullback-Leibler divergence).
Throughout this paper, all logarithms are base-e.
The standard $(n-1)$-simplex is denoted by $\distribution_n$.
The set of all maps from $A$ to $B$ is denoted by the power set $B^A$.
The support set of a map $f$ is denoted by $\support(f)$.
The minimum and the maximum of $x$ and $y$ are denoted by $x\wedge y$ and $x\vee y$, respectively, and $(x)_+\eqdef x\vee 0$.

The contributions of this work are summarized in the following theorems.
Theorems~\ref{kl.ratio} and \ref{kl.ratio.2} give tight lower and upper bounds \eqref{r.range} on the ratio of relative entropies of two probability distributions with respect to a common third one, where the three distributions are collinear in $\distribution_n$.
Theorem~\ref{threshold} determines the exact universal erasure probability threshold and, more generally, the exact universal threshold \eqref{threshold.definition} on the noise measure \eqref{gamma.definition} of an arbitrary side channel.

\begin{theorem}\label{kl.ratio}
Given $\alpha,\beta\in\distribution_n$, we define $z(t)=t\alpha+(1-t)\beta$ for $t\in [0,1]$.
For $0\le a\le 1$ and $0\le c<b\le 1$, we define $u=z(a)$, $v=z(b)$, and $w=z(c)$.
Suppose $\kl{v}{u}=r\kl{w}{u}$, where $\kl{v}{u}$ and $\kl{w}{u}$ are both finite and positive (which implies $\alpha\ne \beta$, $a\ne b$, and $a\ne c$).
Then
\begin{equation}
\frac{1}{\rho(1-a,1-c,1-b)}
< r
< \rho(a,b,c),
\label{r.range}
\end{equation}
where
\begin{equation}
\rho(a,b,c)
\eqdef \begin{cases}
 \frac{\xi_a(b)}{\xi_a(c)} &\text{if $0\le a<1$},\\
 \frac{1-b}{1-c} &\text{otherwise},
 \end{cases}
\label{rho.definition}
\end{equation}
\begin{equation}
\xi_s(t)
\eqdef \zeta_t(t)-\zeta_t(s),
\label{xi.definition}
\end{equation}
\begin{equation}
\zeta_t(s)
\eqdef s+(1-t)\ln(1-s).
\label{zeta.definition}
\end{equation}

Equivalently,
\begin{enumerate}[\renewcommand{\theenumi}{\alph{enumi}}]
\item\label{equivalent.a}
For fixed $r$, $b$, and $c$,
\begin{equation}
a \in I_{1,\downarrow}\cup I_{1,\uparrow12}\cup I_{1,\uparrow21},
\label{a.range}
\end{equation}
\[
I_{1,\downarrow} = (1-\rho_{\downarrow,1-c,1-b}^{-1}(1/r), \rho_{\downarrow,b,c}^{-1}(r))
\]
\[
I_{1,\uparrow12}
= \setlength{\arraycolsep}{0pt}\begin{cases}
 \multicolumn{2}{l}{(\rho_{\uparrow1,b,c}^{-1}(r),1-\rho_{\uparrow2,1-c,1-b}^{-1}(1/r))}\\
 &\text{if $r\ge\frac{\zeta_b(b)}{\zeta_c(c)}$},\\
 [0,1-\rho_{\uparrow2,1-c,1-b}^{-1}(1/r)) &\text{if $\frac{b}{c}<r<\frac{\zeta_b(b)}{\zeta_c(c)}$},\\
 \emptyset &\text{otherwise},
 \end{cases}
\]
\[
I_{1,\uparrow21}
= \setlength{\arraycolsep}{0pt}\begin{cases}
 \multicolumn{2}{l}{(\rho_{\uparrow2,b,c}^{-1}(r), 1-\rho_{\uparrow1,1-c,1-b}^{-1}(1/r))}\\
 &\text{if $0<r\le\frac{\zeta_{1-b}(1-b)}{\zeta_{1-c}(1-c)}$},\\
 (\rho_{\uparrow2,b,c}^{-1}(r),1] &\text{if $\frac{\zeta_{1-b}(1-b)}{\zeta_{1-c}(1-c)}<r<\frac{1-b}{1-c}$},\\
 \emptyset &\text{otherwise},
 \end{cases}
\]
\begin{subequations}\label{rho.piecewise}
\begin{equation}
\rho_{\uparrow1,b,c} = \rho_{\cdot,b,c}|_{[0,c)},
\quad
\rho_{\uparrow2,b,c} = \rho_{\cdot,b,c}|_{[b,1]},
\end{equation}
\begin{equation}
\rho_{\downarrow,b,c} = \rho_{\cdot,b,c}|_{(c,b]},
\end{equation}
\end{subequations}
where $\rho_{\cdot,b,c}$ denotes the function $\rho$ of the first argument (with other arguments fixed).

\item\label{equivalent.b}
For fixed $r$, $a$, and $c$,
\begin{equation}
b \in I_2(r,a,c)\cup I_3(r,a,c),
\label{b.range}
\end{equation}
where
\begin{IEEEeqnarray*}{rCl}
\wholeline{I_2(r,a,c)}\\
\quad &= &\setlength{\arraycolsep}{0pt}\begin{cases}
 \multicolumn{2}{l}{(1-\xi_{1-a,\uparrow}^{-1}(r\xi_{1-a}(1-c)),\xi_{a,\downarrow}^{-1}(r\xi_a(c)))}\\
 \hspace{8em} &\text{if $c<a$ and $r<1$},\\
 \emptyset &\text{otherwise},
\end{cases}
\end{IEEEeqnarray*}
\begin{IEEEeqnarray*}{rCl}
\wholeline{I_3(r,a,c)}\\
\quad &= &\setlength{\arraycolsep}{0pt}\begin{cases}
 \multicolumn{2}{l}{(\xi_{a,\uparrow}^{-1}(r\xi_a(c))\vee c,1-\xi_{1-a,\downarrow}^{-1}(r\xi_{1-a}(1-c)))}\\
 \hspace{5em} &\text{if $r\le\frac{1}{\rho(1-a,1-c,0)}$},\\
 \multicolumn{2}{l}{(\xi_{a,\uparrow}^{-1}(r\xi_a(c))\vee c,1]}\\
 &\text{if $\frac{1}{\rho(1-a,1-c,0)}<r<\rho(a,1,c)$},\\
 \emptyset &\text{otherwise},
\end{cases}
\end{IEEEeqnarray*}
\begin{equation}
\xi_{a,\downarrow} = \xi_a|_{[0,a]},
\quad
\xi_{a,\uparrow} = \xi_a|_{[a,1]}.
\label{xi.piecewise}
\end{equation}

\item\label{equivalent.c}
For fixed $r$, $a$, and $b$,
\begin{equation}
c \in (1-I_2(1/r,1-a,1-b)) \cup (1-I_3(1/r,1-a,1-b)),
\label{c.range}
\end{equation}
where $1-A\eqdef\set{1-x}{x\in A}$.
\end{enumerate}
\end{theorem}

\begin{theorem}\label{kl.ratio.2}
Given $0\le a\le 1$, $0\le c<b\le 1$, and $a\ne b,c$, we define $u=z(a)$, $v=z(b)$, and $w=z(c)$, where $z(t)=t\alpha+(1-t)\beta$ and $\alpha,\beta\in\distribution_n$.
Then
\(
\rho(a,b,c) = \sup_{(\alpha,\beta)\in Q} \frac{\kl{v}{u}}{\kl{w}{u}},
\)
where $Q$ is the set of all pairs $(\alpha,\beta)$ such that $\kl{v}{u}$ and $\kl{w}{u}$ are finite and positive.

In particular, if $\alpha=(1-\delta f(\delta),\delta f(\delta),0,\ldots,0)$ and $\beta=(1-\delta,\delta,0,\ldots,0)$, where $\delta\in (0,1)$, $0\le f(\delta)<1$, and $\lim_{\delta\to 0^+} f(\delta)=0$, then
\(
\lim_{\delta\to 0}\frac{\kl{v}{u}}{\kl{w}{u}} = \rho(a,b,c).
\)
\end{theorem}

\begin{theorem}\label{threshold}
Let $p_{Y|X,S}$ be a memoryless channel with input $X$, output $Y$, and state $S$ distributed according to $p_S$.
The channel state $S$ is known at the decoder, and a noisy state observation $\tilde{S}$, generated by $S$ through side channel $p_{\tilde{S}\mid S}$, is causally available at the encoder.
Here, $X$, $Y$, $S$, $\tilde{S}$ are over finite alphabets $\mathcal{X}=\{0,1\}$, $\mathcal{Y}$, $\mathcal{S}$, and $\tilde{\mathcal{S}}$, respectively.

\begin{enumerate}[\renewcommand{\theenumi}{\alph{enumi}}]
\item\label{threshold.direct}
If
\begin{IEEEeqnarray*}{rCl}
\gamma(p_{\tilde{S}\mid S})
&\eqdef &\sum_{\tilde{s}\in\tilde{\mathcal{S}}} \min_{s\in\mathcal{S}} p_{\tilde{S}\mid S}(\tilde{s}\mid s)\yesnumber\label{gamma.definition}\\
&\ge &T
\eqdef 1-\xi_{\nb^{-1},\uparrow}^{-1}(\xi_{\nb^{-1}}(0))
\approx 0.325176,\yesnumber\IEEEeqnarraynumspace\label{threshold.definition}
\end{IEEEeqnarray*}
then
\begin{equation}
C(p_{Y\mid X,S}, p_S, p_{\tilde{S}\mid S})
= \underline{C}(p_{Y\mid X,S}, p_S),
\label{threshold.eq1}
\end{equation}
where $C(p_{Y\mid X,S}, p_S, p_{\tilde{S}\mid S})$ and $\underline{C}(p_{Y\mid X,S}, p_S)$ denote the capacities of channel $p_{Y\mid X,S}$ with $\tilde{S}$ causally available and unavailable at the encoder, respectively.

\item\label{threshold.converse}
Suppose $\mathcal{Y}=\mathcal{S}=\{0,1\}$ and $\tilde{\mathcal{S}}=\{0,1,2\}$.
The channel $p_{Y\mid X,S}$ with state $S$ is given by
\begin{IEEEeqnarray}{rCl}
p_{Y\mid X,S=0}
&= &\begin{bmatrix}
 p_{Y\mid X,S}(0\mid 0,0) &p_{Y\mid X,S}(1\mid 0,0)\nonumber\\
 p_{Y\mid X,S}(0\mid 1,0) &p_{Y\mid X,S}(1\mid 1,0)
 \end{bmatrix}\\
&= &\begin{bmatrix}
 1 &0\\
 1-\delta &\delta
 \end{bmatrix},
 \IEEEyesnumber\label{channel.achieve.threshold}\IEEEyessubnumber*\\
p_{Y\mid X,S=1}
&= &\begin{bmatrix}
 1-\delta &\delta\\
 1 &0
 \end{bmatrix},\\
p_S
&= &(p_S(0),p_S(1))
= (1-\delta,\delta),
\end{IEEEeqnarray}
where $\delta\in (0,0.5)$.
For any $\iota\in (0,T)$, if
\begin{IEEEeqnarray*}{rCl}
p_{\tilde{S}\mid S}
&= &\begin{bmatrix}
 p_{\tilde{S}\mid S}(0\mid 0) &p_{\tilde{S}\mid S}(1\mid 0) &p_{\tilde{S}\mid S}(2\mid 0)\\
 p_{\tilde{S}\mid S}(0\mid 1) &p_{\tilde{S}\mid S}(1\mid 1) &p_{\tilde{S}\mid S}(2\mid 1)
 \end{bmatrix}\\
&= &\begin{bmatrix}
 1-\epsilon &0 &\epsilon\\
 0 &1-\epsilon &\epsilon
 \end{bmatrix}
\end{IEEEeqnarray*}
with $\epsilon=T-\iota$ (so that $\gamma(p_{\tilde{S}\mid S})=T-\iota$), then
\begin{equation}
C(p_{Y\mid X,S}, p_S, p_{\tilde{S}\mid S})
> \underline{C}(p_{Y\mid X,S}, p_S).
\label{threshold.eq2}
\end{equation}
for sufficiently small $\delta$.
\end{enumerate}
\end{theorem}

\begin{remark}\label{capacity}
The capacity $C(p_{Y\mid X,S}, p_S, p_{\tilde{S}\mid S})$ is given by
\[
C(p_{Y\mid X,S}, p_S, p_{\tilde{S}\mid S})
= \max_{p_U} I(U;Y,S)
= \max_{p_U} I(U;Y\mid S)
\]
(\cite{shannon_channels_1958}, \cite[eq.~(3)]{caire_capacity_1999}), where $U$ is a random variable over $\mathcal{U}=\mathcal{X}^{\tilde{\mathcal{S}}}$ and satisfies
\begin{IEEEeqnarray*}{rCl}
\wholeline{p_{U,X,Y,S,\tilde{S}}(u,x,y,s,\tilde{s})}\\
&= &p_U(u)p_S(s)p_{\tilde{S}\mid S}(\tilde{s}\mid s)1\{x=u(\tilde{s})\}p_{Y\mid X,S}(y\mid x,s)
\yesnumber\IEEEeqnarraynumspace\label{capacity.joint}
\end{IEEEeqnarray*}
and $\support(p_U)\le\min\{(|\mathcal{X}|-1)|\tilde{\mathcal{S}}|+1,|\mathcal{S}||\mathcal{Y}|\}$ (which is optional, see \cite[Theorem~7.2]{el_gamal_network_2011}).
The capacity $\underline{C}(p_{Y\mid X,S}, p_S)$ is given by
\[
\underline{C}(p_{Y\mid X,S}, p_S)
= \max_{p_X} I(X;Y,S)
= \max_{p_X} I(X;Y\mid S)
\]
(\cite[eq.~(7.2)]{el_gamal_network_2011}).
\end{remark}

\begin{figure}
\centering
\includegraphics{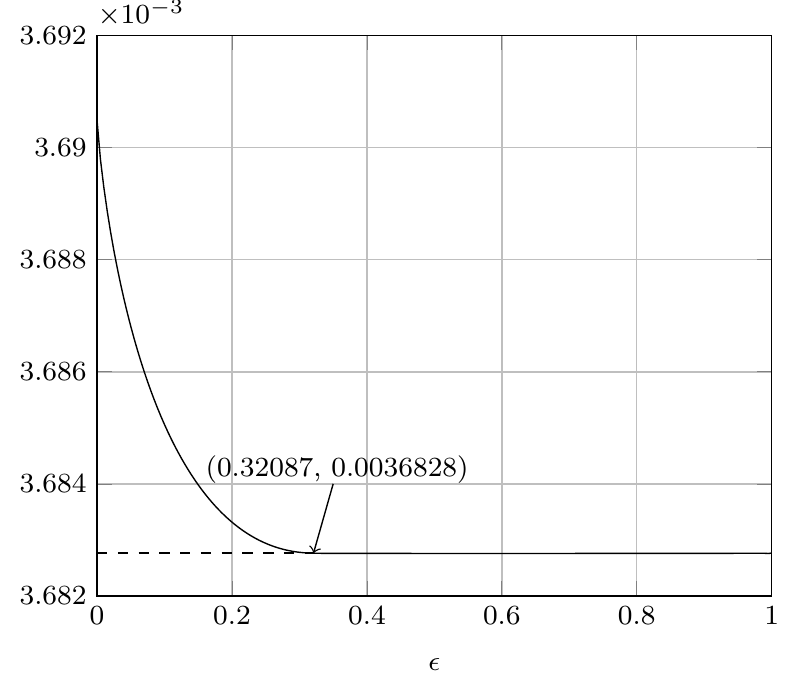}
\caption{A plot of $C(p_{Y\mid X,S}, p_S, p_{\tilde{S}\mid S})$ against $\epsilon$ for $\epsilon\in [0,1]$, where $p_{Y\mid X,S}$ and $p_S$ are given by \eqref{channel.achieve.threshold} with $\delta=0.01$.}
\label{threshold.plot}
\end{figure}

A plot of $C(p_{Y\mid X,S}, p_S, p_{\tilde{S}\mid S})$ against $\epsilon$ for $\epsilon\in [0,1]$ is given in Fig.~\ref{threshold.plot}, where the channel $p_{Y\mid X,S}$ with state $S$ is given by \eqref{channel.achieve.threshold} with $\delta=0.01$.
The erasure probability threshold in this example is very close to the universal threshold $T$ given by \eqref{threshold.definition}.

The rest of this paper is organized as follows.
The proofs of Theorems~\ref{kl.ratio} and \ref{kl.ratio.2} are presented in Section~\ref{proof.kl.ratio}.
The proof of Theorem~\ref{threshold} is given in Section~\ref{proof.threshold}.
Section~\ref{conclusion} contains some concluding remarks.

\section{Proofs of Theorems~\ref{kl.ratio} and \ref{kl.ratio.2}}
\label{proof.kl.ratio}

\begin{IEEEproof}[Proof of Theorem \ref{kl.ratio}]
For any $n$-dimensional probability distribution $p$ such that $p\ll z(1/2)$, we define $f_p(t)=\kl{p}{z(t)}$.
With no loss of generality, we assume that all components of $z(1/2)$ are nonzero.
Then
\(
f_p'(t)
= -\sum_{i=1}^n p_i\frac{\alpha_i-\beta_i}{z_i(t)}
\)
and
\(
f_p''(t)
= \sum_{i=1}^n p_i\frac{(\alpha_i-\beta_i)^2}{(z_i(t))^2},
\)
where $0<t<1$.

The condition $\kl{v}{u}=r\kl{w}{u}$ can be rewritten as
\(
\int_b^a f_v'(t)\diff t
= r\int_c^a f_w'(t)\diff t.
\)
Since for $0<t<1$,
\omitted[{%
\begin{IEEEeqnarray*}{rCl}
\wholeline{(t-b)f_{\alpha}'(t)+(1-t)f_v'(t)}\\
\quad &= &-\sum_{i=1}^n \frac{\alpha_i-\beta_i}{z_i(t)}[t\alpha_i+(1-t)\beta_i](1-b)
= 0
\end{IEEEeqnarray*}
}]{%
\begin{IEEEeqnarray*}{rCl}
\wholeline{(t-b)f_{\alpha}'(t)+(1-t)f_v'(t)}\\
\quad &= &-\sum_{i=1}^n \frac{\alpha_i-\beta_i}{z_i(t)}[(t-b)\alpha_i+(1-t)v_i]\\
&= &-\sum_{i=1}^n \frac{\alpha_i-\beta_i}{z_i(t)}\{(t-b)\alpha_i+(1-t)[b\alpha_i+(1-b)\beta_i]\}\\
&= &-\sum_{i=1}^n \frac{\alpha_i-\beta_i}{z_i(t)}[t\alpha_i+(1-t)\beta_i](1-b)\\
&= &-(1-b)\sum_{i=1}^n (\alpha_i-\beta_i)
= 0
\end{IEEEeqnarray*}
}%
and similarly $(t-c)f_{\alpha}'(t)+(1-t)f_w'(t)=0$, we have
\(
\int_b^a \frac{b-t}{1-t} f_{\alpha}'(t)\diff t
= r\int_c^a \frac{c-t}{1-t} f_{\alpha}'(t)\diff t
\)
or
\begin{equation}
\mathcal{F}_g(r,a,b,c)
= \int_b^a \frac{b-t}{1-t} g(t)\diff t
 - r\int_c^a \frac{c-t}{1-t} g(t)\diff t
= 0,
\label{integration.1}
\end{equation}
where $g(t)=-f_{\alpha}'(t)$.
Since the functions
\(
\frac{b-t}{1-t}
\)
and
\(
\frac{c-t}{1-t}
\)
are not integrable on $(b,1)$ and $(c,1)$, respectively, we assume that $a<1$ and the case of $a=1$ will have to be considered separately.
Since $g'(t)=-f_{\alpha}''(t)$ is negative on $(0,1)$ and
\(
\lim_{t\to 1^-} g(t)\ge \sum_{i=1}^n (\alpha_i-\beta_i) = 0,
\)
$g(t)$ is strictly decreasing and positive on $(0,1)$.
It is also bounded if $g(0)$ is finite.
If however $\lim_{t\to 0} g(t)=+\infty$ (which implies that $\kl{\alpha}{\beta}=+\infty$ and $a\ne 0$), then we define
\[
\tilde{g}(t)
= \begin{cases}
g(t) &\text{if $t\ge \epsilon$},\\
g(\epsilon) &\text{otherwise},
\end{cases}
\]
where $\epsilon$ is a positive number less than all positive numbers in $\{a,b,c\}$.
It follows from \eqref{integration.1} that
\begin{equation}
\mathcal{F}_{\tilde{g}}(r,a,b,c) \le 0
\label{integration.2}
\end{equation}
in all cases, including the case $a>c=0$.
Observing that $g$ or $\tilde{g}$ is positive, bounded, continuous, and strictly decreasing on $(0,1)$, we denote the set of all such functions by $\monotone$.

By \eqref{zeta.definition},
\(
\zeta_t(s)
= \int_0^s \frac{t-t'}{1-t'}\diff t',
\)
so that \eqref{integration.1} with $g(t)=1$ gives
\(
\zeta_b(a)-\zeta_b(b)-r\zeta_c(a)+r\zeta_c(c) = 0.
\)
It is clear that
\(
r^*
= \frac{\zeta_b(b)-\zeta_b(a)}{\zeta_c(c)-\zeta_c(a)}
\)
is the unique solution of this equation, and hence $r < r^*$ (Propositions~\ref{F.property.1} and \ref{F.root}).

In case $a=1$, we have $c<b<a$, so that
\begin{IEEEeqnarray*}{rCl}
\kl{v}{u}
&= &\klbig{\frac{b-c}{a-c}u + \frac{a-b}{a-c}w}{u}
< \frac{1-b}{1-c}\kl{w}{u},
\end{IEEEeqnarray*}
and therefore $s<(1-b)/(1-c)$.

The above arguments prove the second inequality of \eqref{r.range}.
The first inequality can be obtained by exchanging $\alpha$ and $\beta$ with $1-a$, $1-c$, $1-b$, and $1/r$ in place of $a$, $b$, $c$, and $r$, respectively.

We have established the main part of the theorem.
The three equivalent parts are easy consequences of Propositions~\ref{zeta.property}, \ref{rho.property.1}, \ref{rho.property.2}, and \ref{rho.property.3}.
\omitted{%

a) If we fix $r$, $b$, and $c$, then from \eqref{r.range} it follows that
\begin{subequations}\label{equivalent.eq.1}
\begin{equation}
\rho(a,b,c) > r
\end{equation}
\begin{equation}
\rho(1-a,1-c,1-b) > \frac{1}{r},
\end{equation}
\end{subequations}
which combined with Proposition~\ref{rho.property.1} yields
\[
a
\in \rho_{\uparrow1,b,c}^{-1}((r,+\infty)) \cup \rho_{\downarrow,b,c}^{-1}((r,+\infty)) \cup \rho_{\uparrow2,b,c}^{-1}((r,+\infty))
\]
and
\begin{IEEEeqnarray*}{rCl}
a
&\in &1-\rho_{\uparrow1,1-c,1-b}^{-1}((1/r,+\infty))\\
& &\cup\:1-\rho_{\downarrow,1-c,1-b}^{-1}((1/r,+\infty))\\
& &\cup\:1-\rho_{\uparrow2,1-c,1-b}^{-1}((1/r,+\infty)).
\end{IEEEeqnarray*}
Since
\[
\rho_{\uparrow1,b,c}^{-1}((r,+\infty))
= \begin{cases}
 (\rho_{\uparrow1,b,c}^{-1}(r),c) &\text{if $r\ge \frac{\zeta_b(b)}{\zeta_c(c)}$},\\
 [0,c) &\text{otherwise},
\end{cases}
\]
\begin{IEEEeqnarray*}{rCl}
[0,c)
&\supseteq &1-\rho_{\uparrow2,1-c,1-b}^{-1}((1/r,+\infty))\\
&= &\begin{cases}
 [0,1-\rho_{\uparrow2,1-c,1-b}^{-1}(1/r)) &\text{if $r>\frac{b}{c}$},\\
 \emptyset &\text{otherwise},
\end{cases}
\end{IEEEeqnarray*}
and $\zeta_b(b)/\zeta_c(c)>b/c$ (Proposition~\ref{zeta.property}), we have
\begin{equation}
\rho_{\uparrow1,b,c}^{-1}((r,+\infty)) \cap 1-\rho_{\uparrow2,1-c,1-b}^{-1}((1/r,+\infty))
= I_{1,\uparrow12}.
\label{equivalent.eq.a.1}
\end{equation}
Similarly, since
\[
(b,1]
\supseteq \rho_{\uparrow2,b,c}^{-1}((r,+\infty))
= \begin{cases}
(\rho_{\uparrow2,b,c}^{-1}(r),1] &\text{if $r<\frac{1-b}{1-c}$},\\
\emptyset &\text{otherwise},
\end{cases}
\]
\begin{IEEEeqnarray*}{rCl}
(b,1]
&\supseteq &1-\rho_{\uparrow1,1-c,1-b}^{-1}((1/r,+\infty))\\
&= &\begin{cases}
(b,1-\rho_{\uparrow1,1-c,1-b}^{-1}(1/r)) &\text{if $r\le\frac{\zeta_{1-b}(1-b)}{\zeta_{1-c}(1-c)}$},\\
(b,1] &\text{otherwise},
\end{cases}
\end{IEEEeqnarray*}
and $\zeta_{1-b}(1-b)/\zeta_{1-c}(1-c)<(1-b)/(1-c)$ (Proposition~\ref{zeta.property}), we have
\begin{equation}
\rho_{\uparrow2,b,c}^{-1}((r,+\infty)) \cap 1-\rho_{\uparrow1,1-c,1-b}^{-1}((1/r,+\infty))
= I_{1,\uparrow21}.
\label{equivalent.eq.a.2}
\end{equation}
It is also clear that
\begin{IEEEeqnarray*}{rCl}
\wholeline{\rho_{\downarrow,b,c}^{-1}((r,+\infty)) \cap 1-\rho_{\downarrow,1-c,1-b}^{-1}((1/r,+\infty))}\\
\quad &= &(c,\rho_{\downarrow,b,c}^{-1}(r)) \cap (1-\rho_{\downarrow,1-c,1-b}^{-1}(1/r),b)
= I_{1,\downarrow}.
\yesnumber\label{equivalent.eq.a.3}
\end{IEEEeqnarray*}
The equations~\eqref{equivalent.eq.a.1}, \eqref{equivalent.eq.a.2}, and \eqref{equivalent.eq.a.3} together yield \eqref{a.range}.

b) If we fix $r$, $a$, and $c$, then \eqref{equivalent.eq.1} with Propositions~\ref{rho.property.2} and \ref{rho.property.3} yields
\[
b
\in \rho_{a,\downarrow,c}^{-1}((r,+\infty)) \cup \rho_{a,\uparrow,c}^{-1}((r,+\infty))
\]
and
\begin{IEEEeqnarray*}{rCl}
b
&\in &1-\rho_{1-a,1-c,\uparrow}^{-1}((1/r,+\infty))\\
& &\cup\:1-\rho_{1-a,1-c,\downarrow}^{-1}((1/r,+\infty)).
\end{IEEEeqnarray*}
where
\[
\rho_{a,\downarrow,c} = \rho_{a,\cdot,c}|_{(c,a)},
\quad
\rho_{a,\uparrow,c} = \rho_{a,\cdot,c}|_{(a\vee c,1)},
\]
\[
\rho_{a,b,\uparrow} = \rho_{a,b,\cdot}|_{[0,a\wedge b)},
\quad
\rho_{a,b,\downarrow} = \rho_{a,b,\cdot}|_{(a,b)}.
\]
Since
\[
\rho_{a,\downarrow,c}^{-1}((r,+\infty))
= \begin{cases}
 (c,\xi_{a,\downarrow}^{-1}(r\xi_a(c))) &\text{if $c<a$ and $r<1$},\\
 \emptyset &\text{otherwise},
\end{cases}
\]
and
\begin{IEEEeqnarray*}{rCl}
(c,a)
&\supseteq &1-\rho_{1-a,1-c,\downarrow}^{-1}((1/r,+\infty))\\
&= &\setlength{\arraycolsep}{0pt}\begin{cases}
 \multicolumn{2}{l}{(1-\xi_{1-a,\uparrow}^{-1}(r\xi_{1-a}(1-c)),a)}\\
 \hspace{8em} &\text{if $c<a$ and $r<1$},\\
 (c,a) &\text{otherwise},
\end{cases}
\end{IEEEeqnarray*}
we have
\begin{equation}
\rho_{a,\downarrow,c}^{-1}((r,+\infty)) \cap 1-\rho_{1-a,1-c,\downarrow}^{-1}((1/r,+\infty))
= I_2(r,a,c).
\label{equivalent.eq.b.1}
\end{equation}
Since
\begin{IEEEeqnarray*}{rCl}
(a\vee c,1]
&\supseteq &\rho_{a,\uparrow,c}^{-1}((r,+\infty))\\
&= &\begin{cases}
 (\xi_{a,\uparrow}^{-1}(r\xi_a(c))\vee c,1] &\text{if $r<\rho(a,1,c)$},\\
 \emptyset &\text{otherwise},
\end{cases}
\end{IEEEeqnarray*}
\begin{IEEEeqnarray*}{rCl}
(a\vee c,1]
&\supseteq &1-\rho_{1-a,1-c,\uparrow}^{-1}((1/r,+\infty))\\
&= &\setlength{\arraycolsep}{0pt}\begin{cases}
 \multicolumn{2}{l}{(a\vee c,1-\xi_{1-a,\downarrow}^{-1}(r\xi_{1-a}(1-c)))}\\
 \hspace{8em} &\text{if $r\le\frac{1}{\rho(1-a,1-c,0)}$},\\
 (a\vee c,1] &\text{otherwise},
\end{cases}
\end{IEEEeqnarray*}
and $1/\rho(1-a,1-c,0)<\rho(a,1,c)$ implied by \eqref{r.range}, we have
\begin{equation}
\rho_{a,\uparrow,c}^{-1}((r,+\infty)) \cap 1-\rho_{1-a,1-c,\uparrow}^{-1}((1/r,+\infty))
= I_3(r,a,c)
\label{equivalent.eq.b.2}
\end{equation}
Equation \eqref{b.range} then follows from \eqref{equivalent.eq.b.1} and \eqref{equivalent.eq.b.2}.

c) By symmetry, Part~\eqref{equivalent.c} is an easy consequence of Part~\eqref{equivalent.b} with $1/r$, $1-a$, and $1-b$ in place of $r$, $a$, and $c$, respectively.
}
\end{IEEEproof}

\begin{IEEEproof}[{\omitted[Sketch of Proof of Theorem~\ref{kl.ratio.2}]{Proof of Theorem~\ref{kl.ratio.2}}}]
\omitted[%
Use Propositions~\ref{F.root} and \ref{rho.property.1}.%
]{%
Thanks to Theorem~\ref{kl.ratio}, it suffices to prove the second part.
We first assume that $a\ne 1$.
By the proof of Theorem~\ref{kl.ratio},
\begin{IEEEeqnarray*}{rCl}
g(t)
&= &-f_\alpha'(t)\\
&= &\frac{\delta(1-\epsilon)(1-\delta\epsilon)}{(1-\delta\epsilon)t+(1-\delta)(1-t)} - \frac{\delta^2\epsilon(1-\epsilon)}{\delta\epsilon t+\delta(1-t)},
\end{IEEEeqnarray*}
where $\epsilon=f(\delta)$.
It is clear that
\[
g(0)
= \frac{\delta(1-\epsilon)^2}{1-\delta},
\]
so that
\[
|g(0)-\delta(1-\epsilon)|
= \delta(1-\epsilon)\frac{|\delta-\epsilon|}{1-\delta}
\le \delta(1-\epsilon)\frac{\delta\vee\epsilon}{1-\delta}.
\]
Furthermore,
\begin{IEEEeqnarray*}{rCl}
\frac{g(1-\epsilon^{1/2})}{\delta(1-\epsilon)}
&\ge &\frac{1-\delta\epsilon}{1-\delta\epsilon+\epsilon^{1/2}} - \frac{\delta\epsilon}{\delta\epsilon^{1/2}}\\
&= &1 - \epsilon^{1/2} - \frac{\epsilon^{1/2}}{1-\delta\epsilon+\epsilon^{1/2}}
\ge 1 - 3\epsilon^{1/2}
\end{IEEEeqnarray*}
for $\delta$ sufficiently small.
Since $g(t)$ is positive and strictly decreasing,
\begin{IEEEeqnarray*}{rCl}
\wholeline{\int_0^1 |g(t)-\delta(1-\epsilon)|\diff t}\\
&\le &\int_0^{1-\epsilon^{1/2}} \delta(1-\epsilon)M \diff t
 + \int_{1-\epsilon^{1/2}}^1 \delta(1-\epsilon) \diff t\\
&< &\delta(1-\epsilon)(M+\epsilon^{1/2})
\end{IEEEeqnarray*}
for sufficiently small $\delta$, where $M=[(\delta\vee\epsilon)/(1-\delta)]\vee(3\epsilon^{1/2})$.
Then,
\[
\lim_{\delta\to 0} \frac{\norm{g-\delta(1-f(\delta))}_1}{\delta(1-f(\delta))}
= 0,
\]
so that the solution of \eqref{integration.1} (solved for $r$) converges to $\rho(a,b,c)$ as $\delta\to 0$ (Proposition~\ref{F.root}), and therefore $\lim_{\delta\to 0} \kl{v}{u}/\kl{w}{u} = \rho(a,b,c)$.

As for the case $a=1$, we have $0\le c<b<a=1$.
For any $\delta'>0$, since $\kl{v}{u}$ and $\kl{w}{u}$ are positive and finite and $\lim_{a\to 1} \rho(a,b,c)=\rho(1,b,c)$ (Proposition~\ref{rho.property.1}), we let $u'=z(a')$ where $a'$ is arbitrarily close to $1$, so that
\[
\left|\frac{\kl{v}{u}}{\kl{w}{u}}-\frac{\kl{v}{u'}}{\kl{w}{u'}}\right|
\le \delta'
\]
and $|\rho(a',b,c)-\rho(1,b,c)|\le \delta$.
Furthermore, for sufficiently small $\delta$,
\[
\left|\frac{\kl{v}{u'}}{\kl{w}{u'}}-\rho(a',b,c)\right|
\le \delta'.
\]
Then
\[
\left|\frac{\kl{v}{u}}{\kl{w}{u}}-\rho(1,b,c)\right|
\le 3\delta'
\]
for sufficiently small $\delta$.
Since $\delta'$ is arbitrary, the proof is complete.
}%
\end{IEEEproof}

\section{Proof of Theorem~\ref{threshold}}
\label{proof.threshold}

\begin{IEEEproof}
a) To prove \eqref{threshold.eq1}, we need to show that a capacity-achieving input distribution $p_X$ of channel $P_{Y,S\mid X}$ is also optimal for channel $P_{Y,S\mid U}$ (see Remark~\ref{capacity}).

Since $p_X$ is capacity-achieving for $P_{Y,S\mid X}$, it follows from \cite[Theorem~4.5.1]{gallager_information_1968} that
\begin{equation}
\kl{p_{Y,S\mid X=0}}{p_{Y,S}}
= \kl{p_{Y,S\mid X=1}}{p_{Y,S}}
= \underline{C},\label{threshold.proof.eq1}
\end{equation}
where
\begin{equation}
p_{Y,S}(y,s)
= p_S(s) \sum_{x\in\mathcal{X}} p_X(x) p_{Y\mid X,S}(y\mid x,s) \label{threshold.proof.eq2}
\end{equation}
and $\underline{C}=\underline{C}(p_{Y\mid X,S}, p_S)$.
This also implies that
\begin{equation}
p_X(0),p_X(1)
\in (\nb^{-1}, 1-\nb^{-1})
\quad \text{(Theorem~\ref{kl.ratio} with $r=1$)}.\label{threshold.proof.eq3}
\end{equation}
Since $0$ and $1$ can be regarded as constant mappings from $\tilde{\mathcal{S}}$ to $\mathcal{X}$, $p_X$ is also a valid input strategy for $P_{Y,S\mid U}$.
We will show that
\(
\kl{p_{Y,S\mid U=u}}{p_{Y,S}} \le \underline{C}
\)
for all non-constant mappings $u\in\mathcal{U}$, so that the natural (zero) extension of $p_X$ over $\mathcal{U}$, achieves the capacity of $P_{Y,S\mid U}$ (\cite[Theorem~4.5.1]{gallager_information_1968}).

With \eqref{capacity.joint} and \eqref{threshold.proof.eq2}, $\kl{p_{Y,S\mid U=u}}{p_{Y,S}}$ can be expressed as
\omitted[{%
\begin{IEEEeqnarray*}{rCl}
\wholeline{\kl{p_{Y,S\mid U=u}}{p_{Y,S}}}\\
\quad &= &\sum_{y\in\mathcal{Y},s\in\mathcal{S}} p_S(s) p_{Y\mid U,S}(y\mid u,s) \ln\frac{p_{Y\mid U,S}(y\mid u,s)}{p_{Y\mid S}(y\mid s)},
\end{IEEEeqnarray*}
}]{%
\begin{IEEEeqnarray*}{rCl}
\wholeline{\kl{p_{Y,S\mid U=u}}{p_{Y,S}}}\\
\quad &= &\sum_{y\in\mathcal{Y},s\in\mathcal{S}} p_{Y,S\mid U}(y,s\mid u) \ln\frac{p_{Y,S\mid U}(y,s\mid u)}{p_{Y,S}(y,s)}\\
&= &\sum_{y\in\mathcal{Y},s\in\mathcal{S}} p_S(s) p_{Y\mid U,S}(y\mid u,s) \ln\frac{p_S(s) p_{Y\mid U,S}(y\mid u,s)}{p_{Y,S}(y,s)}\\
&= &\sum_{y\in\mathcal{Y},s\in\mathcal{S}} p_S(s) p_{Y\mid U,S}(y\mid u,s) \ln\frac{p_{Y\mid U,S}(y\mid u,s)}{p_{Y\mid S}(y\mid s)},
\end{IEEEeqnarray*}
}%
where
\omitted[{%
\begin{IEEEeqnarray*}{rCl}
\wholeline{p_{Y\mid U,S}(y\mid u,s)}\\
\quad &= &\sum_{x\in\mathcal{X}} P_{Y\mid X,S}(y\mid x,s) \sum_{\tilde{s}\in\tilde{\mathcal{S}}} p_{\tilde{S}\mid S}(\tilde{s}\mid s) 1\{x=u(\tilde{s})\}
\end{IEEEeqnarray*}
}]{%
\begin{IEEEeqnarray*}{rCl}
\wholeline{p_{Y\mid U,S}(y\mid u,s)}\\
\quad &= &\sum_{x\in\mathcal{X}} p_{Y\mid X,S}(y\mid x,s) p_{X\mid U,S}(x\mid u,s)\\
&= &\sum_{x\in\mathcal{X}} P_{Y\mid X,S}(y\mid x,s) \sum_{\tilde{s}\in\tilde{\mathcal{S}}} p_{\tilde{S}\mid S}(\tilde{s}\mid s) 1\{x=u(\tilde{s})\}
\end{IEEEeqnarray*}
}%
and
\begin{equation}
p_{Y\mid S}(y\mid s)
= \sum_{x\in\mathcal{X}} p_X(x) p_{Y\mid X,S}(y\mid x,s).\label{threshold.proof.eq4}
\end{equation}
Then
\begin{IEEEeqnarray*}{rCl}
\wholeline{\kl{p_{Y,S\mid U=u}}{p_{Y,S}}}\\
\quad &= &\sum_{s\in\mathcal{S}} p_S(s) \sum_{y\in\mathcal{Y}} \left( \sum_{x\in\mathcal{X}} p_{X\mid U,S}(x\mid u,s) P_{Y\mid X,S}(y\mid x,s) \right)\\
& &\times\:\ln\frac{\sum_{x\in\mathcal{X}} p_{X\mid U,S}(x\mid u,s) P_{Y\mid X,S}(y\mid x,s)}{\sum_{x\in\mathcal{X}} p_X(x) p_{Y\mid X,S}(y\mid x,s)}
\end{IEEEeqnarray*}
with
\(
p_{X\mid U,S}(x\mid u,s)
= \sum_{\tilde{s}\in\tilde{\mathcal{S}}} p_{\tilde{S}\mid S}(\tilde{s}\mid s) 1\{x=u(\tilde{s})\},
\)
so that $\kl{p_{Y,S\mid U=u}}{p_{Y,S}}$ becomes a function of the channel $p_{X\mid U=u, S}$ from $\mathcal{S}$ to $\mathcal{X}$.
For convenience, we denote this function by $D(\kappa)$ with $\kappa=p_{X\mid U=u, S}$.

By condition, $\gamma(p_{\tilde{S}\mid S})\ge T$, so that $\gamma(p_{X\mid U=u, S})\ge T$ (Proposition~\ref{gamma.product}), and therefore $D(p_{X\mid U=u, S})\le \underline{C}$ (Propositions~\ref{D.property} and \ref{threshold.property} with \eqref{threshold.proof.eq1} and \eqref{threshold.proof.eq3}).

b) \omitted[Use Theorem~\ref{kl.ratio.2} with Propositions~\ref{rho.property.1} and \ref{rho.property.2}.]{%
When state information is not available, we have the channel
\[
p_{Y\mid X}
= \begin{bmatrix}
 1-\delta^2 &\delta^2\\
 1-\delta+\delta^2 &\delta-\delta^2
 \end{bmatrix}
= \begin{bmatrix}
 1-\delta'f(\delta') &\delta'f(\delta')\\
 1-\delta' &\delta'
 \end{bmatrix},
\]
where $\delta'=g(\delta)=\delta(1-\delta)$ which is invertible on $(0,1/2)$, and $f(\delta')=g^{-1}(\delta')/(1-g^{-1}(\delta'))$.

Then it follows from Theorem~\ref{kl.ratio.2} with $\alpha=p_{Y\mid X=0}$, $\beta=p_{Y\mid X=1}$, $a\in(0,1)$, $b=1$, and $c=0$ that
\[
\lim_{\delta\to 0} \frac{\kl{\alpha}{a\alpha+(1-a)\beta}}{\kl{\beta}{a\alpha+(1-a)\beta}}
= \rho(a,1,0)
\]
and $\rho(1-\nb^{-1},1,0)=1$.
Since $\kl{v}{u}/\kl{w}{u}$ is continuous with respect to $a$ and $\rho(t,1,0)$ is strictly decreasing on $(0,1)$ (Proposition~\ref{rho.property.1}), the capacity-achieving input distribution $p_X$ of channel $p_{Y\mid X}$ must satisfy
\[
\lim_{\delta\to 0} \norm{p_X-(1-\nb^{-1},\nb^{-1})}_1 = 0.
\]
On the other hand, it is noticed that for sufficiently small $\delta$,
\[
\kl{p_{Y\mid X=1,S=1}}{p_{Y\mid S=1}}
> \kl{p_{Y\mid X=0,S=1}}{p_{Y\mid S=1}}
\]
with $p_{Y\mid S=1}$ defined by \eqref{threshold.proof.eq4}, so it is tempted to use signal $1$ when $S=1$.
We choose the input strategy
\[
u(\tilde{s})
= \begin{cases}
 1 &\text{if $\tilde{s}=1$},\\
 0 &\text{otherwise}.
 \end{cases}
\]
Because of the random erasure of $p_{\tilde{S}\mid S}$, the actual input distributions under the strategy $u$ are $(1,0)$ and $(\epsilon,1-\epsilon)$ for the states $0$ and $1$, respectively.
By Theorem~\ref{kl.ratio.2} with $\alpha=p_{Y\mid X=1,S=1}$, $\beta=p_{Y\mid X=0,S=1}$, $a=\nb^{-1}$, $b\in (\nb^{-1},1)$, and $c=0$, we have
\[
\lim_{\delta\to 0} \frac{\kl{b\alpha+(1-b)\beta}{a\alpha+(1-a)\beta}}{\kl{\beta}{a\alpha+(1-a)\beta}}
= \rho(\nb^{-1},b,0)
\]
and $\rho(\nb^{-1},1-T,0)=1$.
Since $\kl{v}{u}/\kl{w}{u}$ is continuous with respect to $(a,b)$ (with $0<a<b<1$) and $\rho(\nb^{-1},t,0)$ is strictly increasing on $(\nb^{-1},1)$ (Proposition~\ref{rho.property.2}),
\[
\kl{(1-\epsilon)\alpha+\epsilon\beta}{p_{Y\mid S=1}}
> \kl{\beta}{p_{Y\mid S=1}}
\]
for $\epsilon=T-\iota$ and sufficiently small $\delta$.
Therefore,
\begin{IEEEeqnarray*}{rCl}
\wholeline{\kl{p_{Y,S\mid U=u}}{p_{Y,S}}}\\
&= &D(p_{X\mid U=u,S})\\
&= &p_S(0) D(p_{Y\mid X=0,S=0}\|p_{Y\mid S=0})\\
& &+\: p_S(1) D(\epsilon p_{Y\mid X=0,S=1}+(1-\epsilon) p_{Y\mid X=1,S=1}\|p_{Y\mid S=1})\\
&> &p_S(0) D(p_{Y\mid X=0,S=0}\|p_{Y\mid S=0})\\
& &+\: p_S(1) D(p_{Y\mid X=0,S=1}\|p_{Y\mid S=1})\\
&= &D(\useless_0)
= \kl{p_{Y,S\mid X=0}}{p_{Y,S}},
\end{IEEEeqnarray*}
which implies \eqref{threshold.eq2} (\cite[Theorem~4.5.1]{gallager_information_1968} and Remark~\ref{capacity}), where $D(\cdot)$ is defined in the proof of Part~\eqref{threshold.direct} and $\useless_0$ denotes the deterministic useless channel with constant output $0$.
}
\end{IEEEproof}

\section{Conclusion}
\label{conclusion}

We have established tight lower and upper bounds on the ratio of relative entropies of two probability distributions with respect to a common third one, where the three distributions are collinear in $\distribution_n$ (Theorems~\ref{kl.ratio} and \ref{kl.ratio.2}).
These bounds enable us to settle an open problem left from \cite{xu_when_2017}, namely, determining the exact universal threshold on the noise measure of the side channel (Theorem~\ref{threshold}). 

It is worth noting that \cite[Theorem~2]{shulman_uniform_2004} is a special case of Theorem~\ref{kl.ratio} with $r=1$, $b=1$, and $c=0$.
A natural direction for future work is to extend Theorem~\ref{kl.ratio} to more than two probability distributions and to quantum relative entropy.

\appendices

\section{Properties of $\mathcal{F}_g(r,a,b,c)$}

\begin{proposition}\label{F.property.1}
Let
\begin{equation}
\mathcal{F}_g(r,a,b,c)
\eqdef \int_b^a \frac{b-t}{1-t} g(t)\diff t - r\int_c^a \frac{c-t}{1-t} g(t)\diff t,
\label{F.definition}
\end{equation}
where $g\in\monotone'$, the set of all positive, bounded, continuous, nonincreasing functions on $(0,1)$.
Then $\mathcal{F}_g(r,a,b,c)$ is strictly increasing in $r$ for fixed $a$, $b$, and $c$ with $0\le a,b,c\le 1$ and $a\ne c,1$.
\end{proposition}

\omitted{%
\begin{IEEEproof}
Observe that
\[
\frac{\partial\mathcal{F}_g(r,a,b,c)}{\partial r}
= -\int_c^a \frac{c-t}{1-t} g(t)\diff t,
\]
which is positive whenever $a\ne c$.
\end{IEEEproof}
}

\begin{lemma}\label{integration.min}
Let $f$ and $g$ be bounded measurable functions on $(I,\borel(I))$ and
$\lambda$ the Lebesgue measure on $\real$, where $I=[c,d]$ with $c<d$.
The function $g$ is nonincreasing on $I$.
If for $s\in I$,
\begin{equation}
\int_{[c,s]} f(t)\lambda(\diff t) \ge 0
\label{integration.min.condition}
\end{equation}
with equality iff $s=c$ or $d$, then
\begin{equation}
\int_I f(t)g(t)\lambda(\diff t)\ge 0
\label{integration.min.bound.1}
\end{equation}
with equality iff $g$ is constant on $(c,d)$, and for any $\mu\in\real$,
\begin{equation}
\int_I f(t)g(t)\lambda(\diff t)
\le M \int_I |g(t)-\mu|\lambda(\diff t),
\label{integration.min.bound.2}
\end{equation}
where $M=\sup_{t\in I}|f(t)|$.
\end{lemma}

\omitted{%
\begin{IEEEproof}
Note that, owing to the nonincreasing property of $g$, the two limits
$g(c^+)$ and $g(d^-)$ always exist.
Let $h(t,t')=f(t)1\{0\le t'\le g(t)\}$, which is clearly integrable on
$I\times\real$.
By Fubini's theorem,
\begin{IEEEeqnarray*}{rCl}
\wholeline{\int_I f(t)g(t)\lambda(\diff t)}\\
\quad &= &\int_I \lambda(\diff t) \int_\real f(t) 1\{0\le t'\le g(t)\} \lambda(\diff t')\\
&= &\int_{I\times\real} h\diff(\lambda\times\lambda)\\
&= &\int_\real \lambda(\diff t') \int_I f(t) 1\{0\le t'\le g(t)\} \lambda(\diff t)\\
&= &\int_\real \lambda(\diff t') \int_{J(t')} f(t) \lambda(\diff t)\\
&= &\int_{[g(d^-),g(c^+)]} \lambda(\diff t') \int_{J(t')} f(t) \lambda(\diff t),\yesnumber\label{integration.min.eq1}
\end{IEEEeqnarray*}
where $J(t')=\set{t\in I}{g(t)\ge t'}$ is an interval $[c,t'')$ or
$[c,t'']$ with $t''\in I$, and \eqref{integration.min.eq1} is because
when $t'\notin [g(d^-),g(c^+)]$, $J(t')$ is $\emptyset$,
$\{c\}$, $[c,d)$, or $[c,d]$, so that $\int_{J(t')} f\diff\lambda=0$
by condition~\eqref{integration.min.condition}.

If $g$ is constant on $(c,d)$, then $g(c^+)=g(d^-)$, so that
$\int_I fg\diff\lambda=0$.
On the other hand, if $g$ is not constant on $(c,d)$, then for any $g(d^-)<t'<g(c^+)$, $J(t')=[c,t'')$ or $[c,t'']$ with $c<t''<d$, hence
$\int_{J(t')} f\diff\lambda>0$ for $t'\in (g(d^-),g(c^+))$, and
therefore $\int_I fg\diff\lambda>0$.
This proves \eqref{integration.min.bound.1}, and
\eqref{integration.min.bound.2} is an easy consequence of the following fact:
\begin{IEEEeqnarray*}{rCl}
\int_I f(t)(g(t)-\mu)\lambda(\diff t)
&= &\int_I f(t)g(t)\lambda(\diff t) - \mu\int_I f(t)\lambda(\diff t)\\
&= &\int_I f(t)g(t)\lambda(\diff t).
\end{IEEEeqnarray*}
\end{IEEEproof}
}

\begin{proposition}\label{F.property.2}
Let
\(
r^* = \rho(a,b,c),
\)
where $\rho(a,b,c)$ is defined by \eqref{rho.definition} with $0\le a<1$, $0\le c<b\le 1$, and $a\ne b,c$.
The functional $\mathcal{F}_g(r^*,a,b,c)$ can be written as $\int_{[p_0,p_1]} f(t)g(t)\lambda(\diff t)$ such that
\(
h(s) = \int_{[p_0,s]} f(t)\lambda(\diff t)
\)
is zero at $s=p_1$ and is strictly increasing on $(p_0,p_2)$ and strictly decreasing on $(p_2,p_1)$ for some $p_2\in (p_0,p_1)$, where $p_0=a\wedge c$, $p_1=a\vee b$, and $\lambda$ is the Lebesgue measure on $\real$.
More specifically, we have:

a) If $0\le a<c<b\le 1$, then
\begin{equation}
f(t) = r^*\frac{c-t}{1-t}1\{a\le t\le c\} - \frac{b-t}{1-t}1\{a\le t\le b\}.\label{F.property.2.eq.1}
\end{equation}
It is strictly decreasing on $(a,c)$ and strictly increasing on $(c,b)$, and it is positive on $(a,d)$ and negative on $(d,b)$ for some $d\in (a,c)$.

b) If $0\le c<a<b\le 1$, then
\begin{equation}
f(t) = r^*\frac{t-c}{1-t}1\{c\le t\le a\} - \frac{b-t}{1-t}1\{a\le t\le b\},\label{F.property.2.eq.2}
\end{equation}
It is strictly increasing on $(c,a)$ and $(a,b)$, and it is positive on $(c,a)$ and negative on $(a,b)$.

c) If $0\le c<b<a<1$, then
\begin{equation}
f(t) = r^*\frac{t-c}{1-t}1\{c\le t\le a\} - \frac{t-b}{1-t}1\{b\le t\le a\},\label{F.property.2.eq.3}
\end{equation}
It is strictly increasing on $(c,b)$ and strictly decreasing on $(b,a)$, and it is positive on $(c,d)$ and negative on $(d,a)$ for some $d\in (b,a)$.
\end{proposition}

\omitted{%
\begin{IEEEproof}
It has been shown in the proof of Theorem~\ref{kl.ratio} that $h(p_1)=0$.
Other properties of $h$ are easy consequences of the remaining part.

Equations~\eqref{F.property.2.eq.1}, \eqref{F.property.2.eq.2}, and \eqref{F.property.2.eq.3} are obviously true in the almost-everywhere sense.
It remains to prove the properties of $f$ in the three cases.

a) For $t\in (a,c)$, it follows from Propositions~\ref{zeta.property} and \ref{rho.property.1} that
\begin{IEEEeqnarray*}{rCl}
f'(t)
&= &-r^*\frac{1-c}{(1-t)^2}+\frac{1-b}{(1-t)^2}\\
&< &\frac{-b(1-c)+(1-b)c}{c(1-t)^2}
= -\frac{b-c}{c(1-t)^2}
< 0,
\end{IEEEeqnarray*}
so $f(t)$ is strictly decreasing on $(a,c)$.
For $t\in (c,b)$,
\[
f(t) = -\frac{b-t}{1-t} = \frac{1-b}{1-t}-1,
\]
which is clearly strictly increasing on $(c,b)$.
By Proposition~\ref{rho.property.0},
\[
\lim_{t\in a^+} f(t)
= f(a)
= r^*\frac{c-a}{1-a}-\frac{b-a}{1-a} > 0.
\]
It is also clear that $f(b)=0$.
Therefore, $f(t)$ is positive on $(a,d)$ and negative on $(d,b)$ for some $d\in (a,c)$.

b) When $t\in (c,a)$,
\[
f(t)
= r^*\frac{t-c}{1-t}
= r^*\frac{1-c}{1-t}-r^*
\]
which is strictly increasing.
When $t\in (a,b)$,
\[
f(t) = \frac{1-b}{1-t}-1,
\]
which is also strictly increasing.
Since $\lim_{t\to c^+} f(t)=f(c)=0$ and $\lim_{t\in b^-} f(t)=f(b)=0$, $f(t)$ is positive on $(c,a)$ and negative on $(a,b)$.

c) For $t\in (c,b)$,
\[
f(t)
= r^*\frac{1-c}{1-t}-r^*,
\]
which is strictly increasing.
For $t\in (b,a)$, it follows from Proposition~\ref{rho.property.1} that
\[
f'(t)
= r^*\frac{1-c}{(1-t)^2}-\frac{1-b}{(1-t)^2}
< \frac{(1-b)-(1-b)}{(1-t)^2}
= 0,
\]
so $f(t)$ is strictly decreasing on $(b,a)$.
By Proposition~\ref{rho.property.0},
\begin{IEEEeqnarray*}{rCl}
\lim_{t\to a^-} f(t)
&= &f(a)
= r^*\frac{a-c}{1-a}-\frac{a-b}{1-a}
< 0.
\end{IEEEeqnarray*}
It is also clear that $\lim_{t\to c^+} f(t)=f(c)=0$ and $\lim_{t\to b^+} f(t)=f(b)>0$.
Therefore, $f(t)$ is positive on $(c,d)$ and negative on $(d,a)$ for some $d\in (b,a)$.
\end{IEEEproof}
}

\begin{proposition}\label{F.root}
The equation $\mathcal{F}_g(r,a,b,c)=0$ solved for $r$ has a unique positive solution $q=q(g)$ for $g\in\monotone'$ and fixed $a$, $b$, $c$  with $0\le a<1$, $0\le c<b\le 1$, and $a\ne b,c$.
Then
\(
q(g)
\le q(1)
= \rho(a,b,c)
\)
for all $g\in\monotone$ with equality iff $g$ is constant on $(a\wedge c, a\vee b)$, where $\rho(a,b,c)$ is defined by \eqref{rho.definition}.
If for some positive real $\mu$,
\(
\norm{g-\mu}_1
< \frac{\mu\xi_a(c)(1-a)}{|a-c|},
\)
then
\[
q(g)
\ge q(1)-\frac{M(1-a)\norm{g-\mu}_1}{\mu\xi_a(c)(1-a)-|a-c|\norm{g-\mu}_1},
\]
where $\norm{g}_1=\int_0^1 |g(t)|\diff t$, $\xi_a$ is defined by \eqref{xi.definition}, and $M=M(a,b,c)$ is a certain positive real number.
\end{proposition}

\omitted{%
\begin{IEEEproof}
The existence and uniqueness of $q(g)$ follows from Proposition~\ref{F.property.1} with the facts $\mathcal{F}_g(0,a,b,c)<0$ and $\lim_{r\to +\infty} \mathcal{F}_g(r,a,b,c)=+\infty$.

It is clear that $q(1)$, the solution of $\mathcal{F}_1(r,a,b,c)=0$, is $\rho(a,b,c)$.
From Propositions~\ref{integration.min} and \ref{F.property.2} it follows that
\begin{equation}
0
\le \mathcal{F}_g(q(1),a,b,c)
\le M(a,b,c)\norm{g-\mu}_1.
\label{F.root.eq1}
\end{equation}
The first inequality of \eqref{F.root.eq1} implies that $q(g)\le q(1)$ with equality iff $g$ is constant on $(a\wedge c, a\vee b)$ (Propositions~\ref{F.property.1} and \ref{integration.min}).
On the other hand,
\begin{IEEEeqnarray*}{rCl}
\wholeline{\mathcal{F}_g(q(1),\alpha,\beta,\gamma)}\\
&= &\mathcal{F}_g(q(1),\alpha,\beta,\gamma)-\mathcal{F}_g(q(g),\alpha,\beta,\gamma)\\
&= &(q(1)-q(g))\int_c^a \frac{t-c}{1-t} g(t)\diff t\\
&\ge &(q(1)-q(g)) \left(\int_c^a \frac{t-c}{1-t} \mu\diff t - \int_c^a \frac{t-c}{1-t}|g(t)-\mu|\diff t \right)\\
&\ge &(q(1)-q(g)) \left(\mu\int_c^a \frac{t-c}{1-t} \diff t - \frac{|a-c|}{1-a} \int_0^1 |g(t)-\mu|\diff t \right)\\
&= &(q(1)-q(g)) \left(\mu\xi_a(c)-\frac{|a-c|}{1-a}\norm{g-\mu}_1\right).
\end{IEEEeqnarray*}
This, combined with \eqref{F.root.eq1}, completes the proof.
\end{IEEEproof}
}

\section{Properties of $\zeta_t(s)$, $\xi_s(t)$, and $\rho(a,b,c)$}

\begin{proposition}\label{zeta.property}
For the function $\zeta_t(s)$ defined by
\eqref{zeta.definition},
\(
\zeta_t'(s) = \frac{t-s}{1-s},
\)
so that $\zeta_t$ is strictly increasing on $(0,t)$ and strictly decreasing on $(t,1)$.
Furthermore, we have
\(
\frac{\zeta_b(b)}{\zeta_c(c)} > \frac{b}{c}
\)
for $0<c<b\le 1$.
\end{proposition}

\omitted{%
\begin{IEEEproof}
The first part is obvious, and the second part can be proved by letting $f(t)=\zeta_t(t)/t$ and noting that
\[
f'(t)
= -\frac{t+\ln(1-t)}{t^2}
> -\frac{t-t}{t^2}
= 0
\]
for $0<t<1$.
Also note that this inequality is equivalent to $\rho(0,b,c)>1/\rho(1,1-c,1-b)$ implied by \eqref{r.range}.
\end{IEEEproof}
}

\begin{proposition}\label{xi.property}
For the function $\xi_s(t)$ defined by \eqref{xi.definition} with $0\le s<1$,
\(
\xi_s(0) = \ln\frac{1}{1-s}-s,
\)
\(
\xi_s(s)=0,
\)
and
\(
\xi_s(1) = 1-s.
\)
$\xi_s$ is continuous on $[0,1]$, and it is strictly decreasing on $(0,s)$ and strictly increasing on $(s,1)$.

On the other hand, when $t$ is fixed, $\xi_s(t)$ is strictly decreasing in $s$ for $s\in (0, t)$ and strictly increasing in $s$ for $s\in (t,1)$.

When $s=1-\nb^{-1}$, $\xi_s(0)=\xi_s(1)$, so that for $s\le 1-\nb^{-1}$, $\xi_s(t)=\xi_s(0)$ has a unique solution on $(s,1]$, and for $s\ge 1-\nb^{-1}$, $\xi_s(t)=\xi_s(1)$ has a unique solution on $[0,s)$.
\end{proposition}

\omitted{%
\begin{IEEEproof}
Observe that
\[
\xi_s'(t) = \ln(1-s)-\ln(1-t)
\]
which is negative on $(0,s)$ and positive on $(s,1)$.
Also note that
\[
\frac{\partial\xi_s(t)}{\partial s} = \frac{s-t}{1-s},
\]
which, as a function of $s$, is negative on $(0, t)$ and positive on $(t, 1)$.
These two facts prove the first and the second parts, respectively.
The last part can be easily proved by noting that $\xi_s(0)$ and $\xi_s(1)$ are strictly increasing and decreasing for $s\in [0,1)$, respectively.
\end{IEEEproof}
}

\begin{proposition}\label{xi.property.2}
Let
\(
\xi_{s,\downarrow}=\xi_s|_{[0,s]}
\text{ and }
\xi_{s,\uparrow}=\xi_s|_{[s,1]}.
\)

Let $f_{t,\uparrow}(s)=\xi_{s,\uparrow}^{-1}(\xi_s(t))$, where $s\in (t,d)$, $t\in [0,1)$, and $d$ is the unique solution of $\xi_d(t)=\xi_d(1)$ for $d\in (t, 1)$.
Then $f_{t,\uparrow}(s)$ is strictly increasing in $s$.

Let $f_{t,\downarrow}(s)=\xi_{s,\downarrow}^{-1}(\xi_s(t))$, where $s\in (d,t)$, $t\in (0,1]$, and $d$ is the unique solution of $\xi_d(t)=\xi_d(0)$ for $d\in (0,t)$.
Then $f_{t,\downarrow}(s)$ is strictly increasing in $s$.
\end{proposition}

\omitted{%
\begin{IEEEproof}
This result is a consequence of Proposition~\ref{xi.property}.
The condition $\xi_d(t)=\xi_d(1)$ ensures that
\[
0 = \xi_s(s) < \xi_s(t) < \xi_d(t) = \xi_d(1) < \xi_s(1)
\]
when $s\in (t,d)$, so that $f_{t,\uparrow}(s)$ is well defined.
For $t<s<s'<d$, $\xi_s(t)<\xi_{s'}(t)$, so that $\xi_{s',\uparrow}^{-1}(\xi_{s'}(t)) > \xi_{s',\uparrow}^{-1}(\xi_s(t)) > s'$.
Similarly, $\xi_s(\xi_{s',\uparrow}^{-1} (\xi_s(t))) > \xi_{s'}(\xi_{s',\uparrow}^{-1} (\xi_s(t))) = \xi_s(t)$, so that $\xi_{s',\uparrow}^{-1}(\xi_s(t))>\xi_{s,\uparrow}^{-1}(\xi_s(t))$, and therefore $f_{t,\uparrow}(s)<f_{t,\uparrow}(s')$.

The condition $\xi_d(t)=\xi_d(0)$ ensures that
\[
0 = \xi_s(s) < \xi_s(t) < \xi_d(t) = \xi_d(0) < \xi_s(0)
\]
when $s\in (d,t)$, so that $f_{t,\downarrow}(s)$ is well defined.
For $d<s<s'<t$, $\xi_s(t)>\xi_{s'}(t)$, so that $\xi_{s,\downarrow}^{-1}(\xi_s(t)) < \xi_{s,\downarrow}^{-1}(\xi_{s'}(t)) < s$.
Similarly, $\xi_{s'}(t) = \xi_s(\xi_{s,\downarrow}^{-1}(\xi_{s'}(t))) < \xi_{s'}(\xi_{s,\downarrow}^{-1}(\xi_{s'}(t)))$, so that $\xi_{s,\downarrow}^{-1}(\xi_{s'}(t))<\xi_{s',\downarrow}^{-1}(\xi_{s'}(t))$, and therefore $f_{t,\downarrow}(s)<f_{t,\downarrow}(s')$.
\end{IEEEproof}
}

\begin{proposition}\label{rho.property.0}
Let $\rho(a,b,c)$ be the function defined by \eqref{rho.definition}.
If $0\le a<c<b\le 1$, then
\(
\rho(a,b,c) > \frac{b-a}{c-a}.
\)
If $0\le c<b<a<1$, then
\(
\rho(a,b,c) < \frac{a-b}{a-c}.
\)
\end{proposition}

\omitted{%
\begin{IEEEproof}
If $0\le a<c<b\le 1$, then by Cauchy's mean value theorem and Proposition~\ref{zeta.property},
\[
\rho(a,b,c)
> \frac{\zeta_b(c)-\zeta_b(a)}{\zeta_c(c)-\zeta_c(a)}
= \frac{\zeta_b'(t)}{\zeta_c'(t)}
= \frac{b-t}{c-t}
> \frac{b-a}{c-a},
\]
where $t\in (a,c)$.
Similarly, if $0\le c<b<a<1$, then
\[
\rho(a,b,c)
< \frac{\zeta_b(b)-\zeta_b(a)}{\zeta_c(b)-\zeta_c(a)}
= \frac{\zeta_b'(t)}{\zeta_c'(t)}
= \frac{t-b}{t-c}
< \frac{a-b}{a-c},
\]
where $t\in (b,a)$.
\end{IEEEproof}
}

\begin{proposition}\label{rho.property.1}
Let $f(t)=\rho(t,b,c)$, where $0\le c<b\le 1$.
Then
\(
f(0) = \frac{\zeta_b(b)}{\zeta_c(c)},
\)
\(
f(c) = +\infty,
\)
\(
f(b) = 0,
\)
and
\(
f(1) = \frac{1-b}{1-c}.
\)
The function $f$ is continuous on $[0,c)$ and $(c,1]$, and it is strictly increasing on $(0,c)$ and $(b,1)$ and strictly decreasing on $(c,b)$.
Let
\(
f_{\uparrow1}=f|_{[0,c)},
\)
and
\(
f_\downarrow=f|_{(c,b]},
\)
and
\(
f_{\uparrow2}=f|_{[b,1]}.
\)
Then, for $s>0$,
\[
f_{\uparrow1}^{-1}((s,+\infty))
= \begin{cases}
 (f_{\uparrow1}^{-1}(s),c) &\text{if $s\ge f(0)$},\\
 [0,c) &\text{otherwise},
\end{cases}
\]
\reducevspace
\[
f_\downarrow^{-1}((s,+\infty)) = (c,f_\downarrow^{-1}(s)),
\]
\reducevspace
\[
f_{\uparrow2}^{-1}((s,+\infty))
= \begin{cases}
 (f_{\uparrow2}^{-1}(s),1] &\text{if $s<f(1)$},\\
 \emptyset &\text{otherwise}.
\end{cases}
\]
\end{proposition}

\omitted{%
\begin{IEEEproof}
It is clear that $f$ is continuous on $[0,c)$ and $(c,1)$.
As for $t=1$,
\begin{IEEEeqnarray*}{rCl}
\lim_{t\to 1} f(t)
&= &\lim_{t\to 1} \frac{\zeta_b(b)-\zeta_b(t)}{\zeta_c(c)-\zeta_c(t)}
= \lim_{t\to 1} \frac{\zeta_b'(t)}{\zeta_c'(t)}\\
&= &\lim_{t\to 1} \frac{b-t}{c-t} \qquad \text{(Proposition~\ref{zeta.property})}\\
&= &f(1).
\end{IEEEeqnarray*}

For the remaining part, it suffices to show that $f'(t)$ is positive on $(0,c)\cup (b,1)$ and negative on $(c,b)$.
We have
\[
f'(t)
= \frac{g(t)}{(\zeta_c(c)-\zeta_c(t))^2},
\]
where
\[
g(t)
= \zeta_c'(t)(\zeta_b(b)-\zeta_b(t))-\zeta_b'(t)(\zeta_c(c)-\zeta_c(t)).
\]
If $0<t<c$, then from Proposition~\ref{zeta.property}, it follows
that
\begin{IEEEeqnarray*}{rCl}
g(t)
&> &\zeta_c'(t)(\zeta_b(c)-\zeta_b(t))-\zeta_b'(t)(\zeta_c(c)-\zeta_c(t))\\
&= &\zeta_c'(t) \frac{\zeta_b'(t')(\zeta_c(c)-\zeta_c(t))}{\zeta_c'(t')}-\zeta_b'(t)(\zeta_c(c)-\zeta_c(t)) \yesnumber\IEEEeqnarraynumspace\label{mean.value.1}\\
&= &(\zeta_c(c)-\zeta_c(t)) \left( \zeta_c'(t)\frac{b-t'}{c-t'}-\zeta_b'(t) \right)\\
&> &(\zeta_c(c)-\zeta_c(t)) \left( \zeta_c'(t)\frac{b-t}{c-t}-\zeta_b'(t) \right)
= 0,
\end{IEEEeqnarray*}
where \eqref{mean.value.1} follows from Cauchy's mean value theorem for some $t'\in (t,c)$.
If $c<t<b$, then it follows from Proposition~\ref{zeta.property} that $\zeta_c'(t)<0$ and $\zeta_b'(t)>0$, so that $g(t)<0$.
If $b<t<1$, then it follows from Proposition~\ref{zeta.property} that
\begin{IEEEeqnarray*}{rCl}
g(t)
&> &\zeta_c'(t)(\zeta_b(b)-\zeta_b(t)) - \zeta_b'(t)(\zeta_c(b)-\zeta_c(t))\\
&= &\zeta_c'(t) \frac{\zeta_b'(t')(\zeta_c(b)-\zeta_c(t))} {\zeta_c'(t')} - \zeta_b'(t)(\zeta_c(b)-\zeta_c(t)) \yesnumber\IEEEeqnarraynumspace\label{mean.value.2}\\
&= &(\zeta_c(b)-\zeta_c(t)) \left( \zeta_c'(t)\frac{b-t'}{c-t'}-\zeta_b'(t) \right)\\
&> &(\zeta_c(b)-\zeta_c(t)) \left( \zeta_c'(t)\frac{b-t}{c-t}-\zeta_b'(t) \right)
= 0,
\end{IEEEeqnarray*}
where \eqref{mean.value.2} follows from Cauchy's mean value theorem
for some $t'\in (b,t)$.
\end{IEEEproof}
}

\begin{proposition}\label{rho.property.2}
Let $f(t)=\rho(a,t,c)$, where $0\le a\le 1$, $0\le c<1$, and $a\ne c$.
Then
\[
f(0)
= \begin{cases}
 \frac{\xi_a(0)}{\xi_a(c)} &\text{if $0\le a<1$},\\
 \frac{1}{1-c} &\text{otherwise},
\end{cases},
\]
\(
f(c)=1,
\)
\(
f(a)=0,
\)
and
\(
f(1)=\frac{\xi_a(1)}{\xi_a(c)}.
\)
The function $f$ is continuous on $[0,1]$, and it is strictly decreasing on $(0,a)$ and strictly increasing on $(a,1)$.
Let
\(
f_\downarrow=f|_{(c,a)}
\)
and
\(
f_\uparrow=f|_{(a\vee c,1)}.
\)
Then, for $s>0$,
\[
f_\downarrow^{-1}((s,+\infty))
= \begin{cases}
 (c,\xi_{a,\downarrow}^{-1}(s\xi_a(c))) &\text{if $c<a$ and $s<1$},\\
 \emptyset &\text{otherwise},
\end{cases}
\]
\reducevspace
\[
f_\uparrow^{-1}((s,+\infty))
= \begin{cases}
 (\xi_{a,\uparrow}^{-1}(s\xi_a(c))\vee c,1] &\text{if $s<f(1)$},\\
 \emptyset &\text{otherwise},
\end{cases}
\]
where $\xi_{a,\downarrow}$ and $\xi_{a,\uparrow}$ are defined in Proposition~\ref{xi.property.2}.
\end{proposition}

\omitted{%
\begin{IEEEproof}
Since $f(t)=(1-t)/(1-c)$ for $a=1$, the proposition is clearly true.
As for $a<1$, note that $f(t)=\xi_a(t)/\xi_a(c)$ and use Proposition~\ref{xi.property}.
\end{IEEEproof}
}

\begin{proposition}\label{rho.property.3}
Let $f(t)=\rho(a,b,t)$, where $0\le a\le 1$, $0<b\le 1$, and $a\ne b$.
Then
\[
f(0)
= \begin{cases}
 \frac{\xi_a(b)}{\xi_a(0)} &\text{if $0\le a<1$},\\
 1-b &\text{otherwise},
\end{cases}
\]
\(
f(a)=+\infty,
\)
\(
f(b)=1,
\)
and
\(
f(1)=\frac{\xi_a(b)}{\xi_a(1)}.
\)
The function $f$ is continuous on $[0,a)$ and $(a,1]$, and it is strictly increasing on $(0,a)$ and strictly decreasing on $(a,1)$.
Let
\(
f_\uparrow=f|_{[0,a\wedge b)}
\)
and
\(
f_\downarrow=f|_{(a,b)}.
\)
Then, for $s>0$,
\[
f_\uparrow^{-1}((s,+\infty))
= \begin{cases}
 (\xi_{a,\downarrow}^{-1}(\xi_a(b)/s),a\wedge b) &\text{if $s\ge f(0)$},\\
 [0,a\wedge b) &\text{otherwise},
\end{cases}
\]
\reducevspace
\[
f_\downarrow^{-1}((s,+\infty))
= \begin{cases}
 (a,\xi_{a,\uparrow}^{-1}(\xi_a(b)/s)) &\text{if $b>a$ and $s>1$},\\
 (a,b) &\text{otherwise},
\end{cases}
\]
where $\xi_{a,\downarrow}$ and $\xi_{a,\uparrow}$ are defined in Proposition~\ref{xi.property.2}.
\end{proposition}

\omitted{%
\begin{IEEEproof}
Since $f(t)=(1-b)/(1-t)$ for $a=1$, the proposition is clearly true.
As for $a<1$, note that $f(t)=\xi_a(b)/\xi_a(t)$ and use Proposition~\ref{xi.property}.
\end{IEEEproof}
}

\section{The properties of $\gamma(\kappa)$ and $D(\kappa)$}

\begin{proposition}\label{gamma.decomposition}
Any channel $\kappa:\mathcal{S}\to\mathcal{X}$ can be decomposed into the following form:
\[
\kappa
= \sum_{x\in\mathcal{X}} \lambda_x(\kappa) \left[\gamma(\kappa)\useless_x+(1-\gamma(\kappa))\kappa'\right],
\]
where $\useless_x$ denotes the deterministic useless channel with constant output $x$, and
\begin{IEEEeqnarray*}{rCl}
\gamma(\kappa)
&\eqdef &\sum_{x\in\mathcal{X}} \min_{s\in\mathcal{S}} \kappa(x\mid s)
\in [0,1],\\
\lambda_x(\kappa)
&\eqdef &\begin{cases}
\frac{\min_{s\in\mathcal{S}} \kappa(x\mid s)}{\gamma(\kappa)} &\text{if $\gamma(\kappa)>0$},\\
\frac{1}{|\mathcal{X}|} &\text{otherwise},
\end{cases}\\
\kappa'(x\mid s)
&= &\begin{cases}
\frac{\kappa(x\mid s) - \min_{s'\in\mathcal{S}} \kappa(x\mid s')}{1-\gamma(\kappa)} &\text{if $\gamma(\kappa)<1$},\\
\frac{1}{|\mathcal{X}|} &\text{otherwise}.
\end{cases}
\end{IEEEeqnarray*}
\end{proposition}

\omitted{%
\begin{IEEEproof}[Sketch of proof]
The proof is straightforward and only involves simple algebraic manipulations.
One thing to note is that $\gamma(\kappa) \le \sum_{x\in\mathcal{X}} \kappa(x\mid s) = 1$ where $s$ is arbitrary.
\end{IEEEproof}
}

Since a channel $\kappa:\mathcal{S}\to\mathcal{X}$ can be regarded as a $|\mathcal{S}|\times |\mathcal{X}|$ matrix.
The next property of $\gamma(\kappa)$ is given in a matrix form.

\begin{proposition}\label{gamma.product}
Let $A$ be an $m\times\ell$ channel matrix and $B$ an $\ell\times n$ deterministic channel matrix.
Let $g_j$ be the gap between the least number and the second least number of column $A_{*,j}$ and let $g=\min_{1\le j\le\ell} g_j$.
Then
\(
\gamma(AB)\ge\gamma(A)+(|M|-n)_+g,
\)
where $M=I(\{1,\ldots,\ell\})$ and $I(j)=\arg\min_i A_{i,j}$.
\omitted{%
When $|M|\le n$, the lower bound can be attained by choosing $B$ such
that $|I(B^{-1}(k))|\le 1$ for every $1\le k\le n$, where $B$ is
understood as a map.
(There may be more than one rows attaining the minimum value of
$A_{*,j}$, in which case, it does not matter which row index is
assigned to $I(j)$ because $g=0$).
}
\end{proposition}

\omitted{%
\begin{IEEEproof}
Since $B$ is deterministic,
\[
AB
= C
= \begin{pmatrix}
C_{*,1} &C_{*,2} &\cdots &C_{*,n}
\end{pmatrix}
\]
with every column $C_{*,k}=\sum_{j\in B^{-1}(k)} A_{*,j}$.
Let $I'(k)=\arg\min_i C_{i,k}$.
Then the set $M'=I'(\{1,\ldots,n\})$ has at most $\min\{m,n\}$
elements, and hence misses at least $(|M|-\min\{m,n\})_+$ indices in
$M$, so that at least $(|M|-\min\{m,n\})_+$ columns of $A$ do not
contribute their minimum components to the minimum components of
columns of $C$, and therefore
\begin{IEEEeqnarray*}{rCl}
\gamma(C)
&= &\sum_{k=1}^n C_{I'(k),k}\\
&\ge &\sum_{j=1}^\ell \min_{1\le i\le m} A_{i,j}+(|M|-\min\{m,n\})_+g\\
&= &\gamma(A)+(|M|-n)_+g.
\end{IEEEeqnarray*}
The remaining part is straightforward.
\end{IEEEproof}
}

\begin{proposition}\label{D.property}
Let
\[
D(\kappa)
\eqdef \sum_{s\in\mathcal{S}} p_S(s) \kl{\kappa(\cdot\mid s)\otimes p_{Y\mid X,S=s}}{p_{Y\mid S=s}}
\]
where $\kappa$ is a channel from $\mathcal{S}$ to $\mathcal{X}$,
\[
(\kappa(\cdot\mid s)\otimes p_{Y\mid X,S=s})(y)
= \sum_{x\in\mathcal{X}} \kappa(x\mid s)p_{Y\mid X,S}(y\mid x,s),
\]
and
\(
p_{Y\mid S}(y\mid s)
= \sum_{x\in\mathcal{X}} p_X(x) p_{Y\mid X,S}(y\mid x,s).
\)
If $\mathcal{X}=\{0,1\}$, $D(\useless_0)=D(\useless_1)=C$, and $\gamma(\kappa)\ge T(p_X(0))\vee T(p_X(1))$, then $D(\kappa) \le C$, where
\begin{IEEEeqnarray*}{rCll}
T(a)
&= &\min\{&t\in [0,1]\colon \kl{z(t)}{z(a)} \le \kl{z(1)}{z(a)}\\
& & &\text{for all $\alpha,\beta \in \distribution_{|\mathcal{Y}|}$,}\\
& & &\text{where $z(t)=t\alpha+(1-t)\beta$}\}
\yesnumber\label{threshold.definition.2}
\end{IEEEeqnarray*}
for $a\in (0,1)$.
\end{proposition}

\omitted{%
\begin{IEEEproof}
We denote by $A$ the set on which the minimum is taken in \eqref{threshold.definition.2}.
We will show that it is closed, so that $T(a)$ is well defined.
The set $A$ can be rewritten as
\[
A
= \cap_{\alpha,\beta\in\distribution_{|\mathcal{Y}|}} A(\alpha,\beta)
\]
with $A(\alpha,\beta)=\set{t\in [0,1]}{\kl{z(t)}{z(a)} \le \kl{z(1)}{z(a)}}$.
Since $\kl{z(t)}{z(a)}$, as a function of $t$, is continuous, $A(\alpha,\beta)$ is closed for all $\alpha,\beta$, and hence the intersection $A$ is also closed.

By the convexity of Kullback-Leibler divergence (or the log-sum inequality), it is easy to see that $D(\kappa)$ is convex.
Then,
\[
D(\kappa)
\le \sum_{x\in\mathcal{X}} \lambda_{x} D(\gamma\useless_x+(1-\gamma)\kappa'),
\quad \text{(Proposition~\ref{gamma.decomposition})}
\]
where $\lambda_x = \lambda_x(\kappa)$ and $\gamma=\gamma(\kappa)$.
For every $x\in\mathcal{X}$,
\begin{IEEEeqnarray*}{rCl}
\wholeline{D(\gamma\useless_x+(1-\gamma)\kappa')}\\
\quad &= &\sum_{s\in\mathcal{S}} p_S(s) \mathrm{D}\bigl(\gamma p_{Y\mid X=x,S=s}\\
& &\qquad +\:(1-\gamma)(\kappa'(\cdot\mid s)\otimes p_{Y\mid X,S=s}) \bigm\| p_{Y\mid S=s}\bigr)\\
&= &\sum_{s\in\mathcal{S}} p_S(s) \klbig{\sum_{x'\in\mathcal{X}} \lambda'_x p_{Y\mid X=x',S=s}}{p_{Y\mid S=s}},
\end{IEEEeqnarray*}
where $\lambda_x'\ge\gamma$.
Since $\gamma\ge T(p_X(0))\vee T(p_X(1))$, it follows from \eqref{threshold.definition.2} that $D(\gamma\useless_x+(1-\gamma)\kappa')\le D(\useless_x) = C$.
Therefore, $D(\kappa)\le \sum_{x\in\mathcal{X}} \lambda_x C = C$.
\end{IEEEproof}
}

\begin{proposition}\label{threshold.property}
For $a>\nb^{-1}$, the function $T(a)$ defined by \eqref{threshold.definition.2} can be computed by
\(
T(a) = 1-\xi_{1-a,\uparrow}^{-1}(\xi_{1-a}(0)),
\)
which is strictly increasing in $(\nb^{-1},1)$.
\end{proposition}

\omitted{%
\begin{IEEEproof}
Since $a>\nb^{-1}$, $\xi_{1-a}(0)<\xi_{1-a}(1)$ (Proposition~\ref{xi.property}).
From Theorems~\ref{kl.ratio} and \ref{kl.ratio.2} with $b=1$, it follows that
\begin{IEEEeqnarray*}{rCl}
T(a)
&= &\inf\set{c}{1/\rho(1-a,1-c,0) \ge 1}\\
&= &\inf\set{c}{\rho(1-a,1-c,0) \le 1}\\
&= &\inf (1-[0,\xi_{1-a,\uparrow}^{-1}(\xi_{1-a}(0))]) \yesnumber\label{threshold.property.proof.eq1}\\
&= &\inf [1-\xi_{1-a,\uparrow}^{-1}(\xi_{1-a}(0)),1]
= 1-\xi_{1-a,\uparrow}^{-1}(\xi_{1-a}(0)),
\end{IEEEeqnarray*}
where \eqref{threshold.property.proof.eq1} follows from Proposition~\ref{rho.property.2}.
It is also clear that $T(a)$ is strictly increasing in $(\nb^{-1},1)$ (Proposition~\ref{xi.property.2}).
\end{IEEEproof}
}

\section*{Acknowledgment}

This work was supported in part by the National Natural Science Foundation of China under Grant 61571398 and in part by the Natural Science and Engineering Research Council (NSERC) of Canada under a Discovery Grant.

\bibliographystyle{IEEEtran}
\bibliography{IEEEabrv,ut2}

\end{document}